\documentclass[conference]{IEEEtran}
\IEEEoverridecommandlockouts

\usepackage{graphicx}
\usepackage{mathrsfs,hyperref}
\usepackage{stackengine}
\usepackage{algorithm}
\usepackage{bm}
\usepackage{siunitx}
\usepackage{subcaption}
\usepackage{algpseudocode}\usepackage{xspace}
\usepackage[export]{adjustbox}
\usepackage{xcolor}
\usepackage{makecell}
\usepackage{amsmath,bm,amsfonts}
\usepackage{svg}

\makeatletter
\DeclareRobustCommand\onedot{\futurelet\@let@token\@onedot}
\def\@onedot{\ifx\@let@token.\else.\null\fi\xspace}

\def\eg{\emph{e.g}\onedot} 
\def\ie{\emph{i.e}\onedot} 
 
\def\etc{\emph{etc}\onedot} 
 
\def\etal{\emph{et al}\onedot}
\makeatother

\algnewcommand{\LineComment}[1]{\State \(\triangleright\) #1}

\usepackage[capitalize]{cleveref}
\crefname{section}{Sec.}{Secs.}
\crefname{section}{Section}{Sections}
\crefname{table}{Table}{Tables}
\crefname{table}{Tab.}{Tabs.}
\crefname{algorithm}{Alg.}{Algs.}
\crefname{equation}{}{}

\def\BibTeX{{\rm B\kern-.05em{\sc i\kern-.025em b}\kern-.08em
    T\kern-.1667em\lower.7ex\hbox{E}\kern-.125emX}}

\newcommand{\insertwithsubimagenew}[3][200 120 190 180]
{\stackinset{l}{-0.25cm}{t}{-0.25cm}
  {\scalebox{0.6}{\frame{\includegraphics[trim=#1,clip,width=1.3cm]{#3}}}}
  {\includegraphics[width=#2]{#3}}}

\newcommand{\imagewithinset}[3]
{\stackinset{l}{-0.25cm}{t}{-0.25cm}
  {\scalebox{0.8}{\frame{\includegraphics[clip,width=1.3cm]{#1}}}}
  {\includegraphics[width=#2]{#3}}}
    
\begin{document}

\title{
Let There Be Light: Robust Lensless Imaging Under External Illumination With Deep Learning
\thanks{This work was supported in part by the Swiss National Science
Foundation under Grant CRSII5\textunderscore213521 ``DigiLight---Programmable Third-Harmonic Generation (THG) Microscopy Applied to Advanced Manufacturing''.

Copyright 2025 IEEE. Published in \textit{ICASSP 2025 – 2025 IEEE International Conference on Acoustics, Speech and Signal Processing (ICASSP)}, scheduled for 6-11 April 2025 in Hyderabad, India. Personal use of this material is permitted. However, permission to reprint/republish this material for advertising or promotional purposes or for creating new collective works for resale or redistribution to servers or lists, or to reuse any copyrighted component of this work in other works, must be obtained from the IEEE. Contact: Manager, Copyrights and Permissions / IEEE Service Center / 445 Hoes Lane / P.O. Box 1331 / Piscataway, NJ 08855-1331, USA. Telephone: + Intl. 908-562-3966.
}
}

\author{\IEEEauthorblockN{Eric Bezzam, Stefan Peters, Martin Vetterli}
\IEEEauthorblockA{\textit{Audiovisual Communications Laboratory} \\
\textit{École Polytechnique Fédérale de Lausanne (EPFL)}\\
Lausanne, Switzerland\\
\texttt{first.last@epfl.ch}}
}

\maketitle

\begin{abstract}
Lensless cameras relax the design constraints of traditional cameras by shifting image formation from analog optics to digital post-processing. While new camera designs and applications can be enabled, lensless imaging is very sensitive to unwanted interference (other sources, noise, \etc). In this work, we address a prevalent noise source that has not been studied for lensless imaging: external illumination \eg from ambient and direct lighting. Being robust to a variety of lighting conditions would increase the practicality and adoption of lensless imaging. To this end, we propose multiple recovery approaches that account for external illumination by incorporating its estimate into the image recovery process. At the core is a physics-based reconstruction that combines learnable image recovery and denoisers, all of whose parameters are trained using experimentally gathered data. Compared to standard reconstruction methods, our approach yields significant qualitative and quantitative improvements. We open-source our implementations and a 25K dataset of measurements under multiple lighting conditions.
\end{abstract}

\begin{IEEEkeywords}
lensless imaging, ambient lighting, external illumination, background subtraction, learned reconstruction.
\end{IEEEkeywords}

\section{Introduction}

Cameras are everywhere: in our pockets, in space, and sometimes in our bodies.
We require that cameras work robustly in a variety of contexts to image the far and the seemingly invisible.
Computational lensless imaging disrupts the conventional notions of imaging systems, enabling new designs and novel applications.
By replacing the optics with a thin modulating mask, 
an imaging system can be made compact, low-cost, and provide visual
privacy~\cite{boominathan2022recent}.
Moreover, the multiplexing property of lensless cameras allows for compressive imaging, \eg hyperspectral~\cite{Monakhova:20} and videos~\cite{antipa2019video} from a single capture.
For lensless cameras, a viewable image is not directly formed on the sensor, 
but by a computational algorithm. 
Image recovery is typically posed as a regularized inverse problem on the measurement by physically modeling the imaging system~\cite{flatcam,Antipa:18}.
With a sufficiently large dataset of lensless and ground-truth pairs,
model-based optimization optimization can be combined with neural networks to significantly improve image quality~\cite{Monakhova:19,9239993,Kingshott:22,Li:23}.

While high quality results can be obtained,
previous works have shown the sensitivity of lensless imaging to mismatch in the forward modeling and to input noise variations.
Zeng \etal~\cite{9546648} demonstrate the accumulation of errors due to model mismatch when applying the alternating direction method of multipliers (ADMM) method~\cite{ADMM}.
Moreover, learned methods can break down when there are slight differences in the signal-to-noise ratio (SNR) between training and inference~\cite{Rego2021,Perron2023}.

\begin{figure}[t!]
    \centering
    \includegraphics[width=0.96\linewidth]{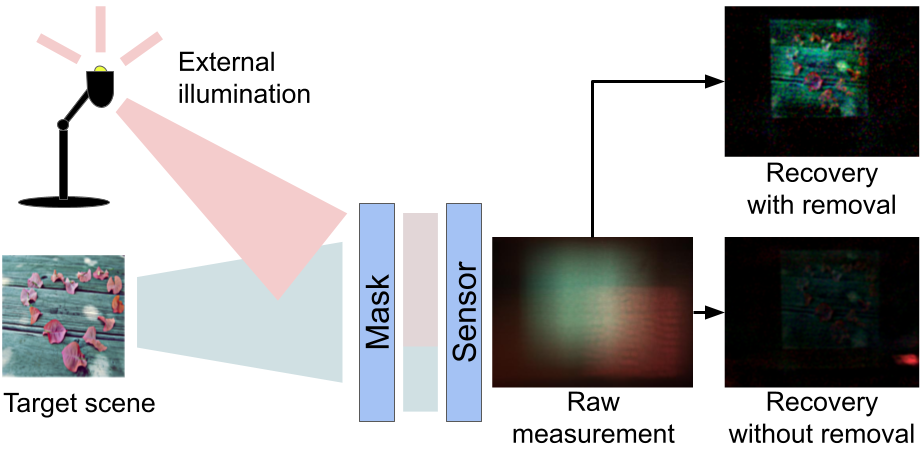}
  \caption{Lensless imaging under external illumination. Recovery is done with the alternating direction method of multipliers (ADMM) method~\cite{ADMM}. Without removing the external illumination through digital post-processing, recovery of a target object may not be discernible (bottom right).}
  \label{fig:external_viz}
  \vspace{-1em}
\end{figure}

One source of noise/interference that has not been studied in lensless imaging is \textit{external illumination}, 
\ie light not emitted from the object of interest.
This can take many forms and is unavoidable in practical settings,
\eg ambient lighting such as natural outdoor lighting or diffuse indoor illumination, and/or direct lighting such as from lamps.
Most works ignore the influence of external illumination by performing measurements in a very controlled, dark environment such that only the object(s) of interest are emitting light:
\eg either displaying images on a screen
or directly shining a light on the object(s)~\cite{Monakhova:19,9239993}.
These controlled environments can be very different from scenarios in which lensless cameras may be deployed,
thereby limiting the practicality of lensless systems if they can only be used in such contexts.
The challenge with external illumination is that the multiplexing property of lensless cameras causes unwanted sources to spread across the entire captured image,
potentially drowning out signal from the object of interest,
as visualized by \cref{fig:external_viz}.

\textbf{Contributions:} In this work, we theoretically show the sensitivity of lensless imaging to external illumination,
and propose methods to address this practical issue.
Concretely, we propose techniques that use an estimate of the external illumination within a physics-based machine learning reconstruction.
Moreover, we open-source the first lensless dataset (25K examples) measured under varied lighting conditions~\cite{ambientdata},
and our reconstruction code in \textit{LenslessPiCam}~\cite{Bezzam2023}.\footnote{\href{https://lensless.readthedocs.io}{lensless.readthedocs.io}}

\section{Background}

\subsection{Modeling of Lensless Imaging}

A lensless imaging system is typically modeled as a linear mapping between a scene of incoherent point sources and a system matrix $\bm{H}$~\cite{Antipa:18}:
\begin{align}
    \label{eq:forward}
    \bm{y} = \bm{H}\bm{x} + \bm{n},
\end{align}
where $\bm{y}$ and $\bm{x}$ are the vectorized lensless measurement and scene intensity respectively, and $\bm{n}$ is additive noise.

A common assumption for simplifying calibration and compute is to assume a shift invariant system, turning \cref{eq:forward} into a 2D convolution with the on-axis point spread function (PSF)~\cite{boominathan2022recent}.
Using the convolution theorem, \cref{eq:forward} can be written as a point-wise multiplication in the frequency domain.
The PSF can be measured, \eg with a white LED at far-field in a dark room, or simulated if the mask structure is known~\cite{9239993,Li:23}.

\subsection{Sensitivity to Model Mismatch}

Image recovery can be posed as an optimization problem:
\begin{align}
\label{eq:opt_gen}
   \bm{\hat{x}} = \arg \min_{\bm{x}} \frac{1}{2} ||\bm{H}\bm{x} - \bm{y}||_2^2 + \lambda \mathcal{R}(\bm{x}),
\end{align}
where $\mathcal{R}(\cdot)$ is a regularization function on the estimate image.
This can be solved with an iterative algorithm~\cite{Monakhova:20,Antipa:18},
Wiener filtering~\cite{flatcam}, or learned approaches~\cite{Monakhova:19,9239993,Li:23}.
However, these methods can be sensitive to mismatch in the system matrix $\bm{H}$,
\eg either the PSF estimate is noisy, or the shift invariant assumption is too simplistic.
If we denote our estimate system matrix as $\bm{\hat{H}}=(\bm{H}+\bm{\Delta}_H)$ where the deviation from the true system matrix is $\bm{\Delta}_H$,
our forward model can be written as:
\begin{align}
    \label{eq:mismatch_forward}
    \bm{y} = \bm{H}\bm{x} + \bm{n} = (\bm{\hat{H}} - \bm{\Delta}_H)\bm{x} + \bm{n}.
\end{align}
Assuming the system is invertible and with spectral radius $\rho(\bm{H}) < 1$, applying direct inversion with $\bm{\hat{H}}$ yields~\cite{9157433}:
\begin{align}
   \label{eq:inversion_terms}
   \bm{\hat{x}} &= \bm{\hat{H}}^{-1} \bm{y} = \bm{\hat{H}}^{-1} (\bm{H}\bm{x} + \bm{n}) \\
   \label{eq:mismatch_breakdown}
   &= \bm{x} - \underbrace{\bm{H}^{-1}\bm{\Delta}_H \bm{x}}_{\text{model mismatch}}\nonumber\\ &\quad + \underbrace{(\bm{I} - \bm{H}^{-1}\bm{\Delta}_H)\bm{H}^{-1}\bm{n}}_{\text{noise amplification}} + \mathcal{O}(\| \bm{\Delta}_H\|_F^2).
\end{align}
A similar breakdown can be shown for ADMM~\cite{9546648}.
While additive error may not fully capture mismatch due the shift invariance assumption, 
the mismatch analysis is simplified without significantly compromising its validity and insight.

\subsection{Robust Lensless Imaging}

Considering the decomposition in \cref{eq:mismatch_breakdown}, robust lensless imaging needs to (1) minimize model mismatch $\bm{\Delta}_H$, (2) minimize input noise $\bm{n}$, and/or (3) reduce the amplification of both.
Model mismatch can be minimized
by learning the PSF~\cite{9239993,Kingshott:22} or performing neural-network transformations of it~\cite{Li:23}.
Input noise can be reduced with a pre-processor
prior to camera inversion~\cite{Perron2023}.
Post-processing can perform perceptual enhancements~\cite{Monakhova:19,9239993}, 
and also reduce the amplified error terms in \cref{eq:mismatch_breakdown}.
For iterative algorithms, the intermediate outputs can be used to reduce the accumulating error~\cite{9546648}.

\section{Methodology}
\label{sec:proposed}

In this section, we demonstrate lensless imaging's sensitivity to external illumination and multiple approaches to address it.
The noise $\bm{n}$ in \cref{eq:forward} can be further decomposed as:
\begin{align}
    \label{eq:noise_breakdown}
    \bm{n} =  \bm{n}_a + \bm{H}\bm{x}_b,
\end{align}
where $\bm{n}_a$ is noise at the sensor, and $\bm{x}_b$ consists of illumination from external objects and ambient lighting that go through the imaging system's mask.
Inserting \cref{eq:noise_breakdown} for $\bm{n}$ in \cref{eq:mismatch_breakdown} yields:
\begin{align}
    \label{eq:direct_inversion_ext}
   \bm{\hat{x}} &= \bm{x} - \underbrace{\bm{H}^{-1}\bm{\Delta}_H \bm{x}}_{\text{model mismatch}} + \underbrace{(\bm{I} - \bm{H}^{-1}\bm{\Delta}_H)\bm{H}^{-1}\bm{n}_a}_{\text{additive noise amplification}}\nonumber\\
   & + \underbrace{(\bm{I} - \bm{H}^{-1}\bm{\Delta}_H)\bm{x}_b}_{\text{external illumination amplification}} + \mathcal{O}(\| \bm{\Delta}_H\|_F^2).
\end{align}

Our objective is to reduce $\bm{\bm{x}_b}$ by making use of an estimate of the external illumination:
\begin{align}
    \label{eq:external_estimate}
    \bm{\hat{b}} = \bm{H}\bm{x}_b - \bm{n}_b,
\end{align}
where $\bm{n}_b$ is the error in the estimate.
The estimate $\bm{\hat{b}}$ can be obtained by
\eg averaging over multiple video frames~\cite{1400815} or taking a measurement when the target object is not present.
Therefore, our approach is best suited for cameras that are installed at fixed locations, \eg for surveillance.


\subsubsection{Direction Subtraction}
The simplest approach for handling external illumination is to directly subtract the estimate prior to image recovery,
\ie $\bm{y} - \bm{\hat{b}} =\bm{H}\bm{x} + \bm{n}_a + \bm{n}_b$, 
such that $(\bm{n}_a + \bm{n}_b)$ is amplified instead of~\cref{eq:noise_breakdown}:
\begin{align}
    \label{eq:direct_inversion_sub}
   \bm{\hat{x}} &= \bm{x} - \underbrace{\bm{H}^{-1}\bm{\Delta}_H \bm{x}}_{\text{model mismatch}} + \underbrace{(\bm{I} - \bm{H}^{-1}\bm{\Delta}_H)\bm{H}^{-1}\bm{n}_a}_{\text{additive noise amplification}}\nonumber\\
   & + \underbrace{(\bm{I} - \bm{H}^{-1}\bm{\Delta}_H)\bm{H}^{-1}\bm{n}_b}_{\text{residual external illumination amplification}} + \mathcal{O}(\| \bm{\Delta}_H\|_F^2).
\end{align}
While \cref{eq:direct_inversion_sub} still has an error term associated with the external illumination,
we expect $\bm{n}_b$ to be smaller than $\bm{H}\bm{x}_b$,
such that the noise amplified by \cref{eq:direct_inversion_sub} is smaller than the noise amplified by \cref{eq:direct_inversion_ext}, \ie $\|\bm{n}_a + \bm{n}_b\| < \|\bm{n}_a + \bm{H}\bm{x}_b\|$.

\subsubsection{Learned Subtraction} 
\begin{figure}[t!]
    \centering
    \includegraphics[width=\linewidth]{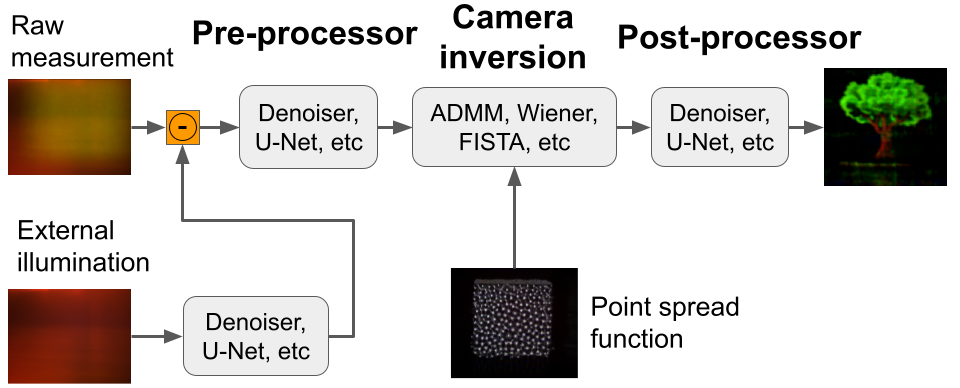}
  \caption{Proposed architecture to address external illumination. Alternatively, the external illumination can be concatenated to the raw measurement to input both to the pre-processor.}
  \label{fig:proposed_pipeline}
  \vspace{-1em}
\end{figure}

To reduce the error $\bm{n}_b$ in the external illumination and thus in \cref{eq:direct_inversion_sub}, we can learn a neural-network transformation of $\bm{\hat{b}}$,
similar to previous approaches that input the PSF to a neural network to reduce model mismatch~\cite{Li:23}.
\cref{fig:proposed_pipeline} shows our proposed architecture for incorporating external illumination removal into the lensless imaging pipeline.
Prior to applying the pre-processor, the processed version of $\bm{\hat{b}}$ is subtracted from the measurement.

\subsubsection{Concatenate External Illumination}

The previous approach introduces a strong inductive bias by using a subtraction as the first (and only) operation between the measurement and the external illumination estimate.
Alternatively, we can concatenate them as a single input to the pre-processor, such that the appropriate operations can be learned between the two.
This is similar to FFDNet~\cite{zhang2018ffdnet} and DRUNet~\cite{zhang2021plug}, which concatenate a single-channel noise level map to the input image.
In our case, we concatenate three-channel external illumination estimate for a six-channel input to the pre-processor.

\section{Experiments}

\subsection{Hardware Setup}

\begin{figure}[t!]
    \centering
    \includegraphics[width=0.7\linewidth]{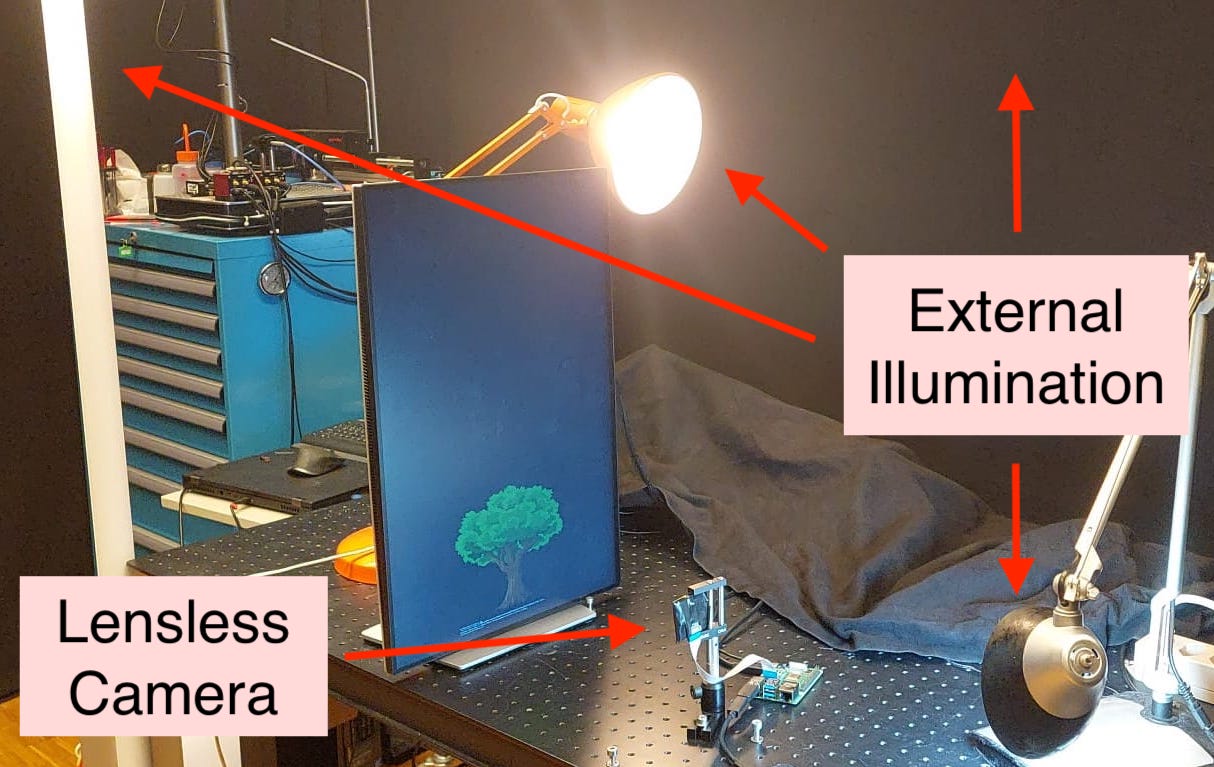}
  \caption{Example setup for measurements under one configuration of external illumination, \ie ceiling lighting and lamps.}
  \label{fig:setup}
  \vspace{-1em}
\end{figure}

\begin{figure}[t!]
    \centering
	\begin{subfigure}{0.39\linewidth}
		\centering
\includegraphics[width=0.99\linewidth]{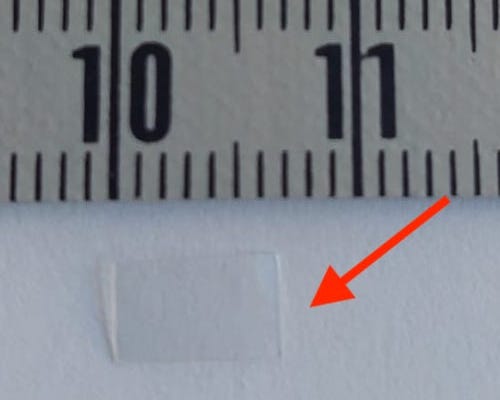}
		\caption{Mask.}
		\label{fig:mask}
	\end{subfigure}
	\begin{subfigure}{0.425\linewidth}
		\centering
    \includegraphics[width=0.99\linewidth]{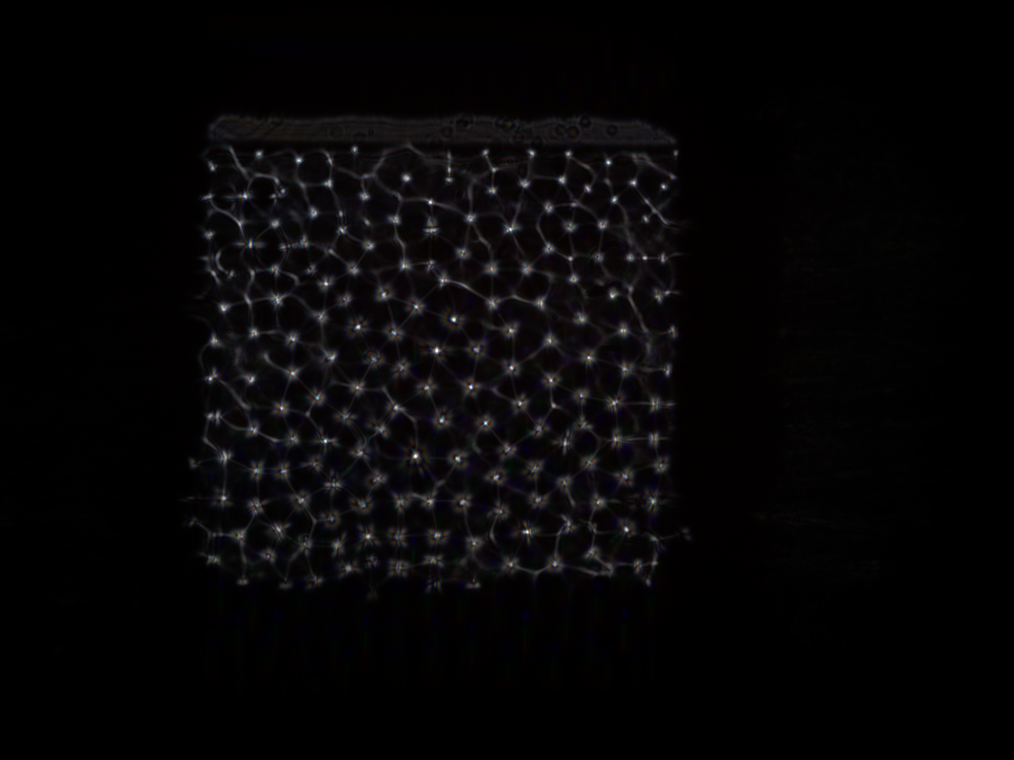}
        \caption{Point spread function.}
		\label{fig:psf}
	\end{subfigure}
	\caption{Mask and point spread function of the lensless camera prototype used in this work, \ie multi-focal mask pattern.}
 \vspace{-1em}
\end{figure}

Our lensless camera and measurement setup can be seen in \cref{fig:setup}.
A monitor is placed \SI{30}{\centi\meter} from the camera to project images,
while our camera consists of a phase mask $\approx\SI{2}{\milli\meter}$ from a Raspberry Pi HQ sensor~\cite{rpi_hq}.
The phase mask (\cref{fig:mask}) has a multi-focal pattern of size $(\SI{3}{\milli\meter}\times\SI{3}{\milli\meter})$ that has been fabricated using ultraviolet single-exposure photolithography~\cite{Lee:23}.
The PSF can be seen in \cref{fig:psf}.

We collect a dataset of 25K images from the MirFlickr dataset~\cite{huiskes2008mir} with a lensless camera and under multiple sources of external illumination:
natural outdoor lighting, indoor (diffuse) ceiling lighting, and directional lamps.
During dataset collection (over 10 days), as well as variations from outdoor lighting, we vary the lamp positions for measurements with different external illuminations:
four different positions in the train set and three in the test set.
For each image, we obtain two measurements:
(1) the target scene on the screen and (2) the screen set to all-black.
The latter serves as our external illumination estimate.
The left-most column of \cref{fig:exp1_compare} shows measurements in our test set with different lighting positions; the inset shows the external illumination estimate.
A train-test split of 85-15 is used: 21.25K files for training and 3.75K files for testing.
The dataset is open-sourced on Hugging Face~\cite{ambientdata}.

\newcommand{\figsizemeas}{0.11}
\newcommand{\figsizebench}{0.10}
\begin{figure*}[t!]
\centering
	\begingroup
	\renewcommand{\arraystretch}{1} 
	\begin{tabular}{cc|cccc|cc}
    \makecell{Measurement} &\makecell{Ground-truth}
    & \makecell{LeADMM\\+$\text{Post}_{8}$~\cite{Monakhova:19}} 
    & \makecell{With learned\\subtraction} 
    & \makecell{$\text{Pre}_{4}$ \& $\text{Post}_{4}$\\\cite{Perron2023}} 
    & \makecell{$\text{Pre}_{4}$ \& $\text{Post}_{4}$\\(concatenate)}
    & \makecell{TrainInv\\+$\text{Post}_{8}$~\cite{9239993}}  
    & \makecell{$\text{Pre}_{4}$ \& $\text{Post}_{4}$\\(concatenate)}\\

    \imagewithinset{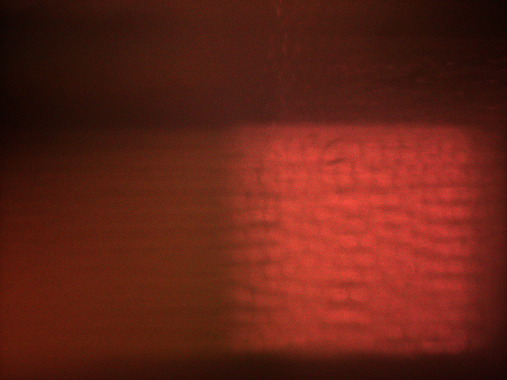}{\figsizemeas\linewidth}{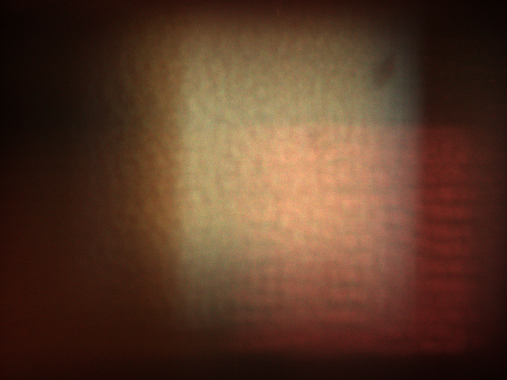}
    &
    \insertwithsubimagenew[45 83 50 8]{\figsizebench\linewidth}{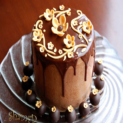} 
    & 
    \insertwithsubimagenew[45 83 50 8]{\figsizebench\linewidth}{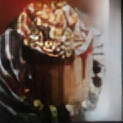}
    & 
    \insertwithsubimagenew[45 83 50 8]{\figsizebench\linewidth}{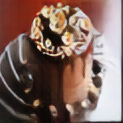}
    & 
    \insertwithsubimagenew[45 83 50 8]{\figsizebench\linewidth}{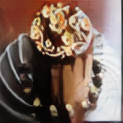}
    & 
    \insertwithsubimagenew[45 83 50 8]{\figsizebench\linewidth}{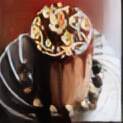}
    & 
    \insertwithsubimagenew[45 83 50 8] 
    {\figsizebench\linewidth}{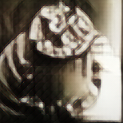} 
    & 
    \insertwithsubimagenew[45 83 50 8]
    {\figsizebench\linewidth}{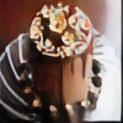} 
    \\

        \imagewithinset{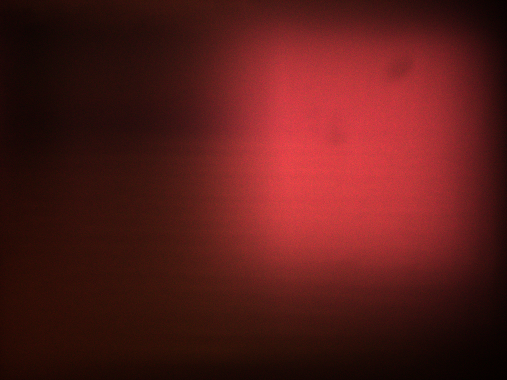}{\figsizemeas\linewidth}{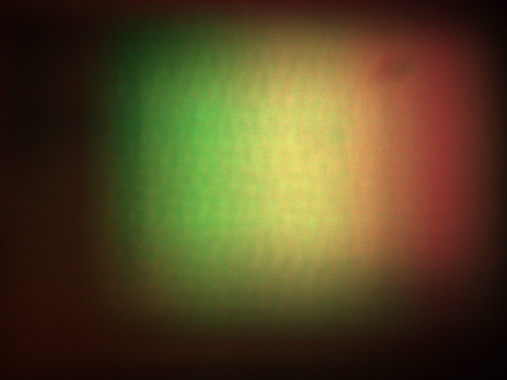}
    &
    \insertwithsubimagenew[85 35 0 45]{\figsizebench\linewidth}{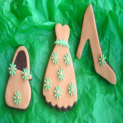} 
    & 
    \insertwithsubimagenew[85 35 0 45]{\figsizebench\linewidth}{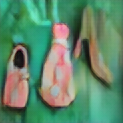}
    & 
    \insertwithsubimagenew[85 35 0 45]{\figsizebench\linewidth}{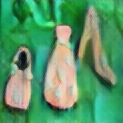}
    & 
    \insertwithsubimagenew[85 35 0 45]{\figsizebench\linewidth}{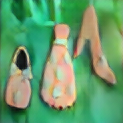}
    & 
    \insertwithsubimagenew[85 35 0 45]{\figsizebench\linewidth}{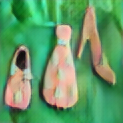}
    & 
    \insertwithsubimagenew[85 35 0 45]
    {\figsizebench\linewidth}{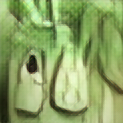} 
    & 
    \insertwithsubimagenew[85 35 0 45]
    {\figsizebench\linewidth}{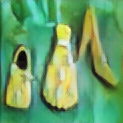} 
    \\

    \imagewithinset{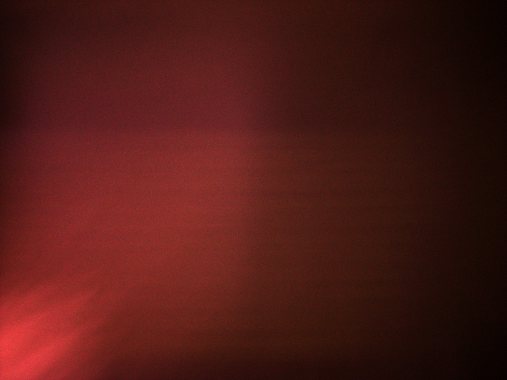}{\figsizemeas\linewidth}{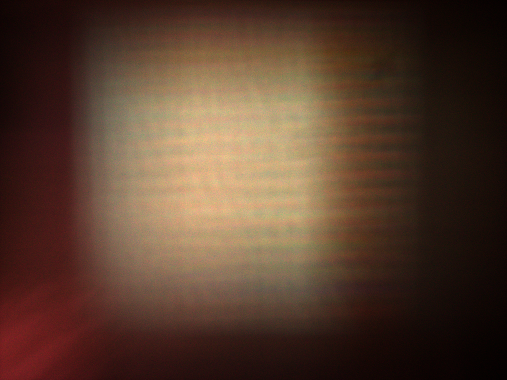}
    &
    \insertwithsubimagenew[50 0 30 70]  
    {\figsizebench\linewidth}{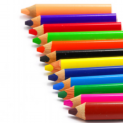} 
    & 
    \insertwithsubimagenew[50 0 30 70]{\figsizebench\linewidth}{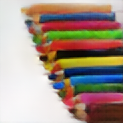}
    & 
    \insertwithsubimagenew[50 0 30 70]{\figsizebench\linewidth}{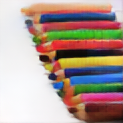}
    & 
    \insertwithsubimagenew[50 0 30 70]{\figsizebench\linewidth}{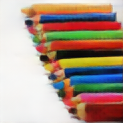}
    & 
    \insertwithsubimagenew[50 0 30 70]{\figsizebench\linewidth}{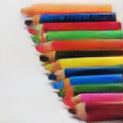}
    & 
    \insertwithsubimagenew[50 0 30 70]
    {\figsizebench\linewidth}{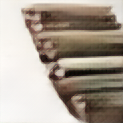} 
    & 
    \insertwithsubimagenew[50 0 30 70]
    {\figsizebench\linewidth}{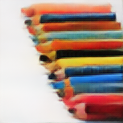} 
    \\
	\end{tabular}
	\endgroup
	\caption{Example reconstruction outputs from the test set of our dataset collected under varied external illumination. Left-most column shows raw measurement with the external illumination in the inset, while the ground-truth is in the second column.}
  \label{fig:exp1_compare}
  \vspace{-1em}
\end{figure*}

\subsection{Reconstruction Approaches}

All approaches can be understood with respect to the modular pipeline in \cref{fig:proposed_pipeline}.
As baselines, we consider two camera inversions approaches: unrolled ADMM with learned parameters (\textit{LeADMM})~\cite{Monakhova:19} and fine-tuning the PSF for single-step inversion (\textit{TrainInv})~\cite{9239993}.
With these camera inverters we either use a neural-network post-processor~\cite{Monakhova:19,9239993}, or both a pre- and post-processor~\cite{Perron2023}.
To the baselines, we incorporate our proposed techniques from \cref{sec:proposed} to address external illumination.
For a valid comparison, we limit the total number of neural network parameters to around 8.1M.
To this end, all denoisers use the DRUNet architecture~\cite{zhang2021plug} with the number of intermediate channels set for a target number of parameters denoted in the subscript, \eg $\textit{Pre}_4$ refers to a pre-processor with around 4M parameters.
For learned subtraction, a DRUNet with just 128K parameters is used for the external illumination and the pre-processor size is slightly reduced.

\subsection{Training and Evaluation}

PyTorch~\cite{Paszke2017} is used for training and evaluation on an Intel Xeon E5-2680 v3 \SI{2.5}{\giga\hertz} CPU
and 4$\times$ Nvidia Titan X Pascal GPUs. 
Training is done with the AdamW optimizer~\cite{loshchilov2017decoupled} 
with $\beta_1=0.9$, $\beta_2=0.999$, $\epsilon=10^{-8}$, and weight decay of $0.01$.
A cosine decay learning rate schedule is used with an initial learning rate of $10^{-4}$, after a warm-up of \SI{5}{\percent} of the training steps (for $25$ epochs and a batch size of $4$).
Our loss is a sum of the mean-squared error and the learned perceptual image patch similarity (LPIPS) with VGG weights~\cite{zhang2018perceptual} between the reconstruction $\bm{\hat{x}}$ and the ground-truth $\bm{x}$:
\begin{equation}
    \label{eq:loss_mse_lpips}
    \mathscr{L}\left(\bm{x},\bm{\hat{x}}\right) = \mathscr{L}_{\text{MSE}}\left(\bm{x},\bm{\hat{x}}\right) + \mathscr{L}_{\text{LPIPS}}\left(\bm{x},\bm{\hat{x}}\right).
\end{equation}
Three metrics are used to evaluate image recovery quality: peak signal-to-noise ratio (PSNR), structural similarity index measure (SSIM), and LPIPS.

\subsection{Results}

When using \textit{LeADMM}~\cite{Monakhova:19} for camera inversion (\cref{tab1} and middle columns of \cref{fig:exp1_compare}),
only using a post-processor results in blurring and loss of detail.
Adding learned subtraction restores finer details 
and improves the metrics: \SI{2}{\decibel} increase and \SI{15}{\percent} and \SI{13}{\percent} relative improvement in SSIM and LPIPS.
Using a pre-processor 
further improves performance, even without accounting for external illumination.
The best performance is obtained with a pre-processor and the proposed subtraction/concatenation techniques.
Using \textit{TrainInv}~\cite{9239993} for camera inversion also benefits from the proposed techniques (\cref{traininv} and last columns of \cref{fig:exp1_compare}). 
While worse than \textit{LeADMM}, it has faster inference~\cite{Perron2023}.

\cref{fig:exp2_direct} shows reconstructions of real objects to show performance beyond images on a screen. 
Our approach improves contrast and is more robust to changes in lighting.

\begin{table}[t!]
\caption{Average image quality metrics on test set with \textit{LeADMM} for camera inversion.}
\begin{center}
\begin{tabular}{|c|c|c|c|}
\hline
\textbf{Method} & \textbf{PSNR} $\uparrow$ & \textbf{SSIM} $\uparrow$ & \textbf{LPIPS}  $\downarrow$ \\
    \hline
    LeADMM+$\text{Post}_{8}$~\cite{Monakhova:19}   & 17.5  & 0.501  & 0.425 \\
     \hline
    \makecell{With direct subtraction}
      & 18.6  & 0.525 & 0.380 \\
    \hline
    With learned subtraction   &  19.5 & 0.577 & 0.368 \\
    \hline
    \hline
    $\text{Pre}_{4}$+LeADMM+$\text{Post}_{4}$~\cite{Perron2023}   & 19.9  &  0.600 & 0.352 \\
         \hline
    \makecell{With direct subtraction}
       & 19.9  & 0.592 & 0.336  \\
    \hline
    With learned subtraction   & 20.5  & 0.618 & 0.331 \\
    \hline
    Concatenate
       &  \textbf{20.6} & \textbf{0.623}  & \textbf{0.329} \\
    \hline
\end{tabular}
\label{tab1}
\end{center}
\vspace{-1em}
\end{table}

\begin{table}[t!]
\caption{Average image quality metrics on test set with \textit{TrainInv} for camera inversion.}
\begin{center}
\begin{tabular}{|c|c|c|c|}
\hline
\textbf{Method} & \textbf{PSNR} $\uparrow$ & \textbf{SSIM} $\uparrow$ & \textbf{LPIPS}  $\downarrow$ \\
\hline
    TrainInv+$\text{Post}_{8}$~\cite{9239993}   & 16.8  & 0.488  & 0.485 \\
     \hline
    \makecell{With direct subtraction}
      & 16.9 & 0.451 & 0.472 \\
    \hline
    With learned subtraction   & 18.0  & 0.535 & 0.445 \\
    \hline
    \hline
    $\text{Pre}_{4}$+TrainInv+$\text{Post}_{4}$~\cite{Perron2023}   & 18.9 & 0.574  &  0.396 \\
         \hline
    \makecell{With direct subtraction}
       & 18.5 & 0.533 & 0.394  \\
    \hline
    With learned subtraction   & 19.5  & 0.560 & 0.410 \\
    \hline
    Concatenate
       & \textbf{20.3}  &  \textbf{0.624} & \textbf{0.355} \\
    \hline
\end{tabular}
\label{traininv}
\end{center}
\vspace{-2em}
\end{table}

\newcommand{\figsizemeasdirect}{0.2}
\newcommand{\figsizerecondirect}{0.19}
\begin{figure}[t!]
\centering
	\begingroup
	\renewcommand{\arraystretch}{1} 
	\setlength{\tabcolsep}{0.3em} 
	\begin{tabular}{ccccc}
    \makecell{Measurement} 
    & \makecell{ADMM100\\(direct sub)} 
    & \makecell{LeADMM\\+$\text{Post}_{8}$~\cite{Monakhova:19}
    } 
    & 
    \makecell{$\text{Pre}_{4}$ \& $\text{Post}_{4}$\\(concatenate)}
    \\

    \imagewithinset{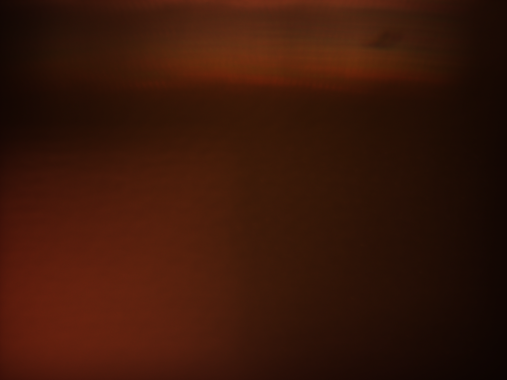}{\figsizemeasdirect\linewidth}{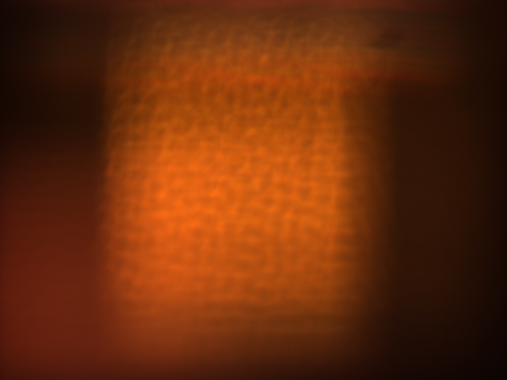}
    &
    \includegraphics[width=\figsizerecondirect\linewidth]{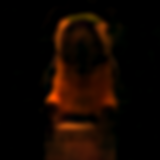}
    & \includegraphics[width=\figsizerecondirect\linewidth]{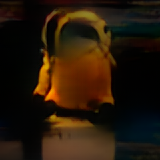}
    & 
\includegraphics[width=\figsizerecondirect\linewidth]{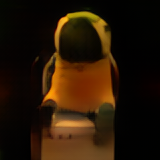}
    \\

    \imagewithinset{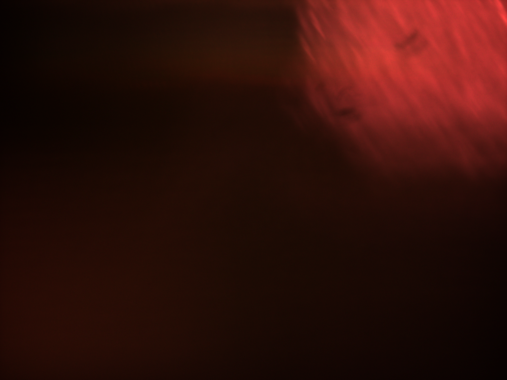}{\figsizemeasdirect\linewidth}{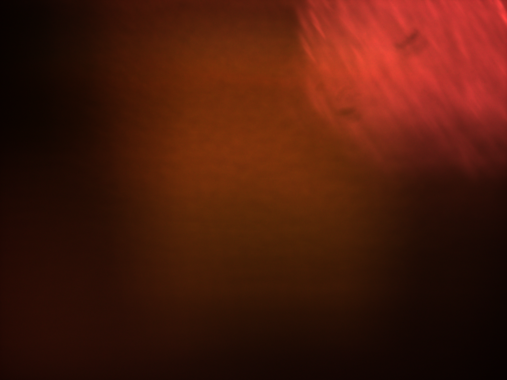}
    &
    \includegraphics[width=\figsizerecondirect\linewidth]{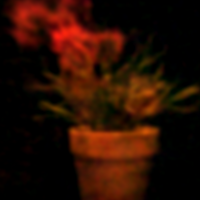}
    & \includegraphics[width=\figsizerecondirect\linewidth]{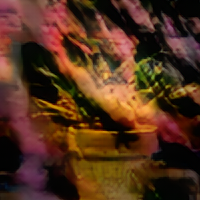}
    & 
    \includegraphics[width=\figsizerecondirect\linewidth]{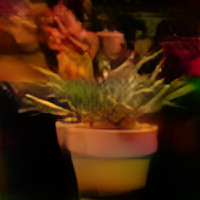}\\
    
    \imagewithinset{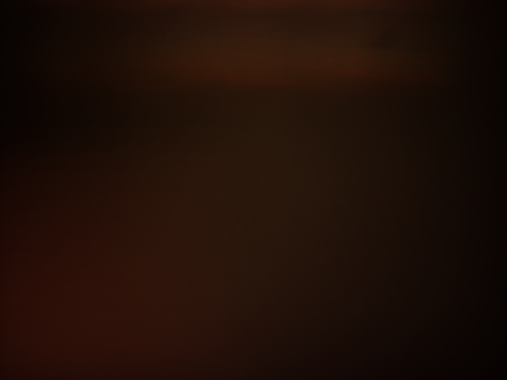}{\figsizemeasdirect\linewidth}{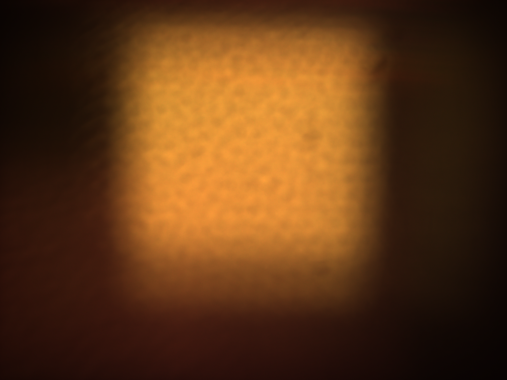}
    &
    \includegraphics[width=\figsizerecondirect\linewidth]{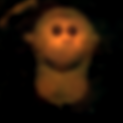}
    & \includegraphics[width=\figsizerecondirect\linewidth]{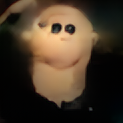}
    & 
    \includegraphics[width=\figsizerecondirect\linewidth]{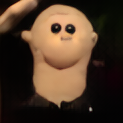}\\

	\end{tabular}
	\endgroup
	\caption{Direct-capture recovery of real objects (plush toys and plant). External illumination is in the inset of left-most column.}
  \label{fig:exp2_direct}
  \vspace{-1.5em}
\end{figure}

\section{Conclusion}

We propose simple yet robust techniques for enabling lensless imaging under external illumination, improving the practicality of such systems beyond controlled setups.
We theoretically show the adverse effects of external illumination to motivate our approaches, 
in which we subtract/concatenate an estimate of the illumination.
We open source our implementations and a dataset measured in varied lighting conditions.
One limitation is the assumption that the target object is not present when the external illumination is estimated.
Source separation techniques~\cite{naik2011overview} could be explore for scenarios where the target and external illumination need to be simultaneously recovered.



\section*{Acknowledgment}

The authors thank Kyung Chul Lee 
for providing the fabricated mask for the lensless camera prototype.


\bibliographystyle{IEEEbib}
\bibliography{main}

\end{document}